\documentclass[letterpaper]{article} 
\usepackage{aaai2026}  
\usepackage{times}  
\usepackage{helvet}  
\usepackage{courier}  
\usepackage[hyphens]{url}  
\usepackage{graphicx} 
\usepackage{natbib}  
\usepackage{caption} 
\usepackage{algorithm}
\usepackage{algorithmic}

\usepackage{newfloat}
\usepackage{listings}
\DeclareCaptionStyle{ruled}{labelfont=normalfont,labelsep=colon,strut=off} 
\floatstyle{ruled}
\newfloat{listing}{tb}{lst}{}
\floatname{listing}{Listing}
\usepackage{booktabs}
\usepackage{tabularray}
\usepackage{multirow}
\usepackage{tabularx}
\usepackage{makecell}
\usepackage{graphicx}
\usepackage{arydshln} 
\usepackage{tabularray}
\UseTblrLibrary{booktabs} 
\usepackage{longtable}  
\usepackage{times}
\usepackage{helvet}
\usepackage{courier}
\usepackage{xcolor}
\usepackage[capitalize]{cleveref}
\usepackage{appendix}

\title{Advancing the Foundation Model for Music Understanding}
\author{
    Yi Jiang\textsuperscript{\rm 1}, Wei Wang\textsuperscript{\rm 2}\thanks{{ }Correspondence to: shakespere@zju.edu.cn}, Xianwen Guo\textsuperscript{\rm 2}, Huiyun Liu\textsuperscript{\rm 2}, Hanrui Wang\textsuperscript{\rm 2},\\ Youri Xu\textsuperscript{\rm 2}, Haoqi Gu\textsuperscript{\rm 2}, Zhongqian Xie\textsuperscript{\rm 2}, Chuanjiang Luo\textsuperscript{\rm 2}
}
\affiliations{
    \textsuperscript{\rm 1}Zhejiang University \quad
    \textsuperscript{\rm 2}NetEase Cloud Music

}

\usepackage{bibentry}

\begin{document}

\maketitle

\begin{abstract}
The field of Music Information Retrieval (MIR) is fragmented, with specialized models excelling at isolated tasks. In this work, we challenge this paradigm by introducing a unified foundation model named MuFun for holistic music understanding. Our model features a novel architecture that jointly processes instrumental and lyrical content, and is trained on a large-scale dataset covering diverse tasks such as genre classification, music tagging, and question answering. To facilitate robust evaluation, we also propose a new benchmark for multi-faceted music understanding called MuCUE (Music Comprehensive Understanding Evaluation). Experiments show our model significantly outperforms existing audio large language models across the MuCUE tasks, demonstrating its state-of-the-art effectiveness and generalization ability.
\end{abstract}

\section{Introduction}
\label{sec:introduction}

Music, a universal and multifaceted form of human expression, presents a formidable challenge for computational understanding. The field of Music Information Retrieval (MIR) has made significant strides in decoding its complex structures. However, progress has historically been characterized by a paradigm of fragmentation. Models are typically engineered as highly specialized experts, excelling at isolated tasks such as genre classification, beat tracking, or instrument recognition. While effective in their narrow domains, this specialization comes at a cost: a lack of holistic, integrative understanding that mirrors human cognition.

This fragmentation engenders significant limitations. Expert models struggle to generalize across tasks and often fail when confronted with complex, multi-faceted queries that require synergistic reasoning. For instance, answering "Why does this song evoke a sense of melancholy?" necessitates a concurrent analysis of its harmony, tempo, instrumentation, and lyrical content—a capability beyond the scope of single-task models. While recent general-purpose audio-language models possess impressive cross-modal capabilities, they are not intrinsically optimized for the unique structural and semantic nuances of music, often lacking the domain-specific acuity for fine-grained MIR tasks.

To bridge this gap and break the prevailing paradigm, we introduce a unified foundation model for holistic music understanding. Our model features an architecture capable of concurrently processing both instrumental audio and lyrical content, and is trained on a vast and diverse corpus of data spanning a multitude of MIR tasks. Instead of cultivating a collection of disparate specialists, our objective is to build a single, versatile generalist model that learns a shared, rich representation of music, enabling it to perform a wide array of tasks from a single set of weights.

A parallel challenge in the pursuit of holistic music understanding lies in its evaluation. The absence of a unified, comprehensive benchmark makes it difficult to compare models systematically or to measure true progress towards general musical intelligence. To address this critical need, we propose the Music Comprehensive Understanding Evaluation (MuCUE) benchmark. MuCUE's core innovation is its standardized format, framing a wide spectrum of tasks—from low-level perception (e.g., pitch and chord recognition) to high-level cognition (e.g., mood and structural analysis)—as multiple-choice questions (MCQs). This approach not only facilitates objective and scalable evaluation but also provides a rigorous tool for probing the emergent reasoning abilities of foundation models, thereby guiding future research.

To summarize, our main contributions are as follows:

\begin{itemize}
    \item \textbf{A Unified Foundation Model for Music:} We introduce a novel, end-to-end trainable foundation model that holistically understands music by jointly processing instrumental audio and lyrical content. Its architecture and a specialized long-context (390s) training regimen enable it to move beyond single-task limitations and achieve state-of-the-art performance on a wide array of MIR tasks.
    \item \textbf{The MuCUE Benchmark:} We propose the Music Comprehensive Understanding Evaluation (MuCUE), a new and extensive benchmark designed to systematically assess music AI capabilities. By framing diverse tasks from low-level perception to high-level cognition as multiple-choice questions, MuCUE provides a standardized and rigorous tool for measuring true progress in the field.
\end{itemize}

\section{Related Work}
\label{sec:related_work}

Recent advancements in music-language understanding leverage frozen audio encoders (often MERT\cite{li2023mert}) integrated with large language models (LLMs) via lightweight adapters to overcome music-text data scarcity. 
MU-LLaMA\cite{liu2023music} (built on LLaMA\cite{touvron2023llama}) pioneered this approach using audio-adapted LLaMA layers and the MusicQA dataset (synthesized from captions and tags), demonstrating strong QA and captioning performance. 
MusiLingo\cite{deng2024musilingo} refined this paradigm by aligning frozen MERT embeddings with LLMs like Vicuna\cite{vicuna2023} through a simple linear projector; its key contribution is the high-quality MusicInstruct (MI) dataset for instruction-tuning, enabling robust open-ended QA and outperforming MU-LLaMA. 
Similarly, LLARK\cite{gardner2023llark} employs an adapter-based architecture trained on augmented data to excel at instruction-following tasks, including detailed captioning and musical reasoning. 
Expanding beyond understanding, M2UGen\cite{liu2023m} introduces a unified LLaMA 2-based framework combining comprehension (music QA, captioning) with cross-modal generation (text/image/video-to-music, editing), utilizing large synthetic instruction datasets (MUCaps, MUEdit etc.) and LoRA fine-tuning to achieve SOTA across both understanding and creative tasks.

While these adapter-based models advance music-text alignment, significant limitations in understanding persist. Crucially, reliance on synthetic or limited datasets (e.g., MusicQA, MI derived primarily from MusicCaps\cite{agostinelli2023musiclm}, MagnaTagATune\cite{law2009evaluation}) restricts scope and depth, potentially perpetuating biases and superficial connections. Scaling QA to complex, subjective musical concepts (emotion, structure, cultural context) remains challenging, as most datasets focus on factual tags or short descriptive captions. Evaluation inadequacies are pronounced: reliance on NLP metrics like BLEU or ROUGE poorly captures musical nuance, subjective meaning, or aesthetic relevance, especially for open-ended QA and long-form captioning. 

There are also much breakthroughs of general audio large language models recently\cite{Qwen2-Audio,abouelenin2025phi,xu2025qwen2,kimiteam2025kimiaudiotechnicalreport,huang2025stepaudiounifiedunderstandinggeneration,zeng2024glm4,fu2025vita,li2025baichuan}.
For instance, Qwen2-Audio\cite{Qwen2-Audio} establishes a high standard as an audio-language model by integrating a Whisper-large-v3\cite{radford2023robust} encoder with a Qwen-7B\cite{qwen} LLM through a refined three-stage training pipeline (pre-training, SFT, DPO) to achieve state-of-the-art performance. 
Expanding the scope beyond audio-language pairs, Qwen2.5-Omni\cite{xu2025qwen2} operates as a truly end-to-end multimodal system processing text, image, audio, and video, distinguished by a "Thinker-Talker" framework for real-time streaming speech generation and a novel positional embedding (TMRoPE) for synchronizing audio-visual inputs. 
Concurrently, Kimi-Audio\cite{kimiteam2025kimiaudiotechnicalreport} proposes a universal audio foundation model, built on a hybrid architecture and pre-trained on over 13 million hours of diverse audio, aiming to unify perception, reasoning, and generation within a single, open-source framework. 
Although these models are not explicitly trained for Music Information Retrieval (MIR), their state-of-the-art speech recognition capabilities provide a strong foundation for interpreting lyrical music. Consequently, their proficiency in processing lyrical content enables them to effectively perform ancillary MIR tasks such as lyrics transcription and, by extension, music genre classification or mood detection where lyrical themes are indicative.

A significant research gap emerges from this dichotomy, defining two distinct frontiers in music-language modeling. On one hand, adapter-based models are purpose-built for music but are often constrained by their architectural design, such as frozen audio backbones and shallow adapters, limiting their capacity for deep semantic representation. They can be characterized as specialized yet brittle. On the other hand, general-purpose multi-modal systems possess powerful, end-to-end trained architectures but lack the domain-specific optimization required to interpret the complex, non-linguistic syntax of music, such as harmony, structure, and expressive nuance. These models are powerful yet unspecialized for nuanced MIR tasks. Our work is strategically positioned at the confluence of these two paradigms, aiming to synergize the architectural power of modern multi-modal systems with the deep, domain-specific focus essential for MIR. The key differentiators of our approach are summarized in Table \ref{tab:comparison}. Notably, our model benefits from full-parameter tuning across all components and, more critically, is trained on an extended audio context of up to 390 seconds—an order of magnitude greater than that of prior models. This strategic combination of a state-of-the-art foundation, comprehensive fine-tuning, and a novel long-context training regimen is designed to create a model that is both broadly capable and musically astute.

\begin{table*}[]
    \resizebox{\textwidth}{!}{%
\begin{tabular}{@{}llllllll@{}}
\toprule
                          & MU-LLaMA        & MusiLingo & LLARK           & M2UGen              & Qwen2-Audio & Kimi-Audio & ours     \\ \midrule
LLM                       & LLaMA-2 7B      & Vicuna-7B & Llama2-7b       & Llama2-7b           & Qwen-7B     & Qwen2.5-7B & Qwen3-8B \\
audio encoder             & MERT            & MERT      & Jukebox         & MERT                & whisper     & whisper    & whisper  \\
audio token frequency(Hz) & fixed embedding & 0.17      & 10              & fixed embedding     & 25          & 12.5       & 10       \\
max train duration(s)     & 29              & 30        & 25              & 10                  & 30(likely)  & 30(likely) & 390      \\
tune modules              & adapter         & adapter   & projection, LLM & adapters, LLM(lora) & unknown     & all        & all      \\ \bottomrule
\end{tabular}%
}
\caption{Comparison of some Audio Language Models(ALMs)}
\label{tab:comparison}
    \end{table*}

\begin{figure*}[t] 
    \centering
    \includegraphics[width=\textwidth]{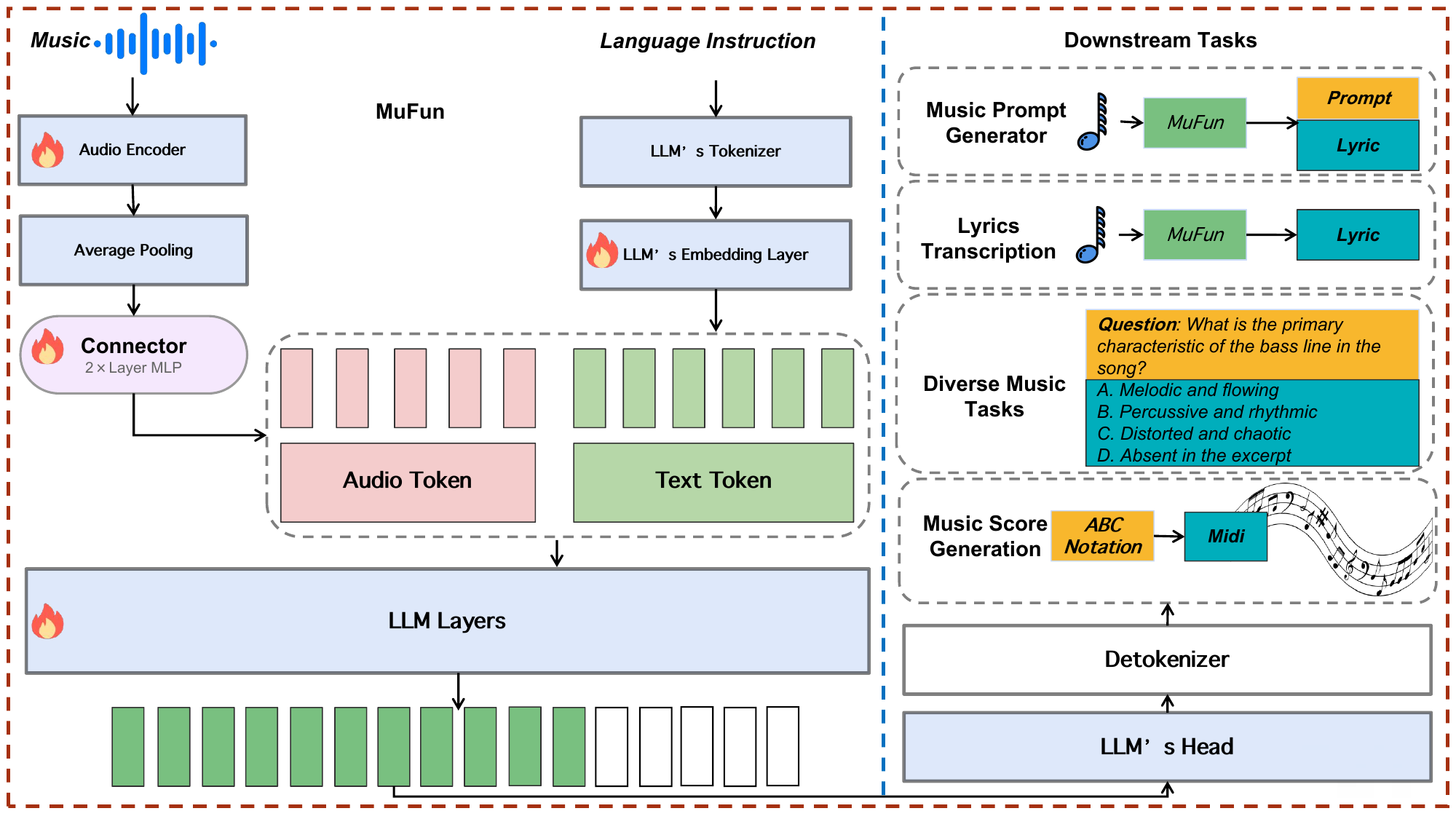} 
    \caption{Overview of the MuFun model architecture
    }
    \label{fig:arch}
\end{figure*}
\label{sec:training}

\section{Model Architecture}
\label{sec:model_architecture}

The architecture of our model is grounded in the effective and scalable designs of recent multimodal large language models. This paradigm, which couples a powerful pre-trained audio encoder with a large language model backbone, provides a robust foundation for complex reasoning tasks. Our model is designed to accept an interleaved sequence of audio and text inputs and generate a coherent text output. Formally, for any input sequence of the form $[A_1, T_1, A_2, T_2, \ldots, A_n, T_n]$, where $A_i$ represents an audio file and $T_i$ a text segment, each modality is first transformed into a sequence of embedding vectors. These embedding sequences are then concatenated and fed into the language model to produce the final output. The overall architecture, depicted in Figure \ref{fig:arch}, comprises three core components: a language model backbone, an audio encoder, and a connector module to bridge the two modalities.

\subsection{Language Model Backbone}
Our language model backbone is initialized from Qwen3-8B-Base \cite{yang2025qwen3}. We selected this model for its state-of-the-art performance in language understanding and multilingual capabilities. Its strong foundational skills are crucial for interpreting the complex, often abstract, relationships within music and for generating nuanced, descriptive text. By leveraging such a powerful pre-trained LLM, we can focus our efforts on effectively translating musical information into a "language" that the LLM can comprehend.
\begin{table*}[]
\centering
\caption{Training Configuration across Different Stages of Model Development}
\label{tab:training_stages}
\resizebox{\textwidth}{!}{%
\begin{tabular}{@{}l|cccc|cc@{}}
\toprule
& \multicolumn{4}{c|}{\textbf{Pretraining}} & \multicolumn{2}{c}{\textbf{Finetuning}} \\
\midrule
\textbf{Sub-stages} & \textbf{Warmup} & \textbf{Align1} & \textbf{Align2} & \textbf{Context Extending} & \textbf{Short Music} & \textbf{Long Music} \\
\midrule
Training steps & 400 & 2500 & 3500 & 540 & 1000 & 700 \\
\midrule
Batch size & 384 & 384 & 384 & 144 & 384 & 144 \\
\midrule
Tune modules & connector & all & all & all & all & all \\
\midrule
Tasks & \makecell[l]{speech transcription;\\ music score transcription} & \makecell[l]{speech transcription;\\ music score transcription} & \makecell[l]{speech transcription;\\ music score transcription;\\ lyrics transcription;\\ pitch, instrument identification} & \makecell[l]{music score transcription;\\ lyrics transcription} & \makecell[l]{various MIR tasks\\ (short audio or segments)} & \makecell[l]{various MIR tasks\\ (mainly song level)} \\
\midrule
Max audio duration (s) & 30 & 30 & 30 & 390 & 30 & 390 \\
\midrule
GPU hours & 10 & 180 & 244 & 175 & 68 & 180 \\
\bottomrule
\end{tabular}%
}
\end{table*}

\begin{figure*}[t] 
    \centering
    \includegraphics[width=\textwidth]{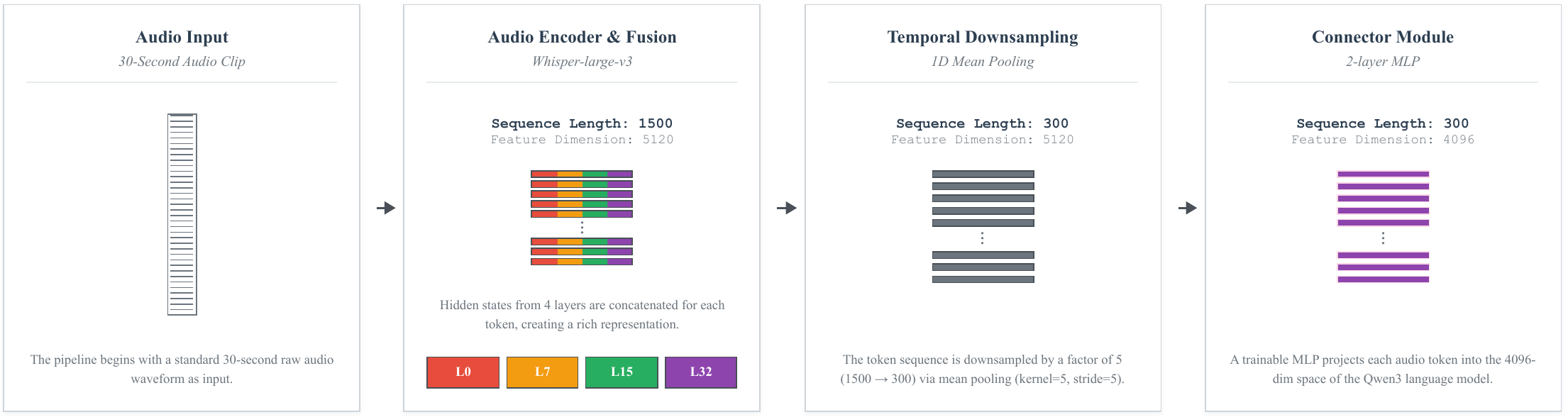} 
    \caption{Pipeline of the Audio processing Module
    }
    \label{fig:audio}
\end{figure*}
\label{sec:training}

\subsection{Audio Encoding and Multi-Layer Feature Fusion}
The audio processing module is responsible for converting raw audio waveforms into meaningful feature representations. For this, we initialize the encoder from Whisper-large-v3 \cite{radford2023robust} as our audio backbone. The motivation for this choice is twofold: Whisper is pre-trained on an enormous and diverse dataset of audio, enabling it to learn a highly robust and general-purpose representation of acoustic phenomena that is transferable to music. Furthermore, its transformer-based architecture is well-suited for capturing temporal dependencies.

To create a comprehensive representation of the audio, we do not rely solely on the final output layer of the Whisper encoder. Instead, we adopt a multi-layer feature fusion strategy. Specifically, we extract the hidden states from four distinct layers of the encoder—layers 0, 7, 15, and 32—and concatenate them. This results in a rich feature vector with a dimension of $5120$ ($1280 \times 4$). The rationale behind this approach is that different layers of a deep network capture different levels of abstraction. Early layers (e.g., layer 0) tend to preserve low-level acoustic details like timbre and pitch, while deeper layers (e.g., layer 32) capture more abstract, semantic information like melodic contours and rhythmic patterns. By providing the model with this multi-resolution view, we empower it to access both fine-grained textural details and high-level structural information simultaneously, a critical requirement for holistic music understanding.

\subsection{Temporal Downsampling and Long-Context Handling}
The Whisper encoder processes a 30-second audio clip into a sequence of 1500 embedding vectors, corresponding to a temporal frequency of 50 Hz. This high density of tokens can be computationally burdensome for the LLM and may not align well with the typical information density of text. To address this, we apply a temporal downsampling step. We use a 1D mean pooling layer with a kernel size and stride of 5 along the time dimension. This operation reduces the audio token frequency to a more manageable 10 Hz, achieving two goals: 1) it significantly reduces the sequence length, improving computational efficiency, and 2) it smooths the representation, encouraging the model to focus on more salient temporal events.

A key ability of our model is to process long-form, song-level audio. To handle inputs exceeding the 30-second window of the Whisper encoder, we employ a straightforward yet effective chunking strategy. The long audio stream is first segmented into 30-second non-overlapping chunks. Each chunk is processed independently by the audio encoder and pooling layer. The resulting embedding sequences are then concatenated in their original order to form a single, continuous sequence representing the entire audio piece. This mechanism extends the model's effective receptive field to any audio duration, enabling true song-level analysis.

\subsection{The Connector Module}
The final architectural component is the connector, which serves as a bridge between the audio and language modalities. Its purpose is to project the 5120-dimensional audio embeddings into the 4096-dimensional space of the Qwen3 language model. For this, we use a 2-layer Multilayer Perceptron (MLP). The MLP first expands the input dimension by a factor of two, applies a GELU non-linear activation function, and then projects it down to the target dimension of the LLM. Using a non-linear MLP instead of a simple linear projection affords greater expressive power\cite{liu2024improved}, allowing for a more complex and nuanced alignment between the learned representations of music and language. This trainable "translator" is vital for harmonizing the two modalities effectively.

\section{Training}
\label{sec:training}
The development of our model's comprehensive musical understanding is facilitated by a meticulously designed, multi-stage training regimen. This protocol is not a monolithic process but rather a strategic curriculum designed to progressively build capabilities. We employ a curriculum learning approach, systematically advancing from foundational audio-text alignment to sophisticated, long-context musical reasoning. The complexity of the tasks and the length of the audio context are gradually increased, ensuring a stable and efficient learning trajectory. The entire training protocol, summarized in Table \ref{tab:training_stages}, is divided into two primary phases: a four-stage pre-training phase to build a robust foundation, and a dual-track fine-tuning phase to specialize the model for diverse MIR applications.

We construct our training data (primarily from public available datasets) in a straight-forward way, unlike many prior works which mostly utilize instruction format. We tend to put as many text labels as possible in a single sample for better efficiency, for which details and some samples are provided in the appendix. Experiments are conducted using NVIDIA A100 40 GB GPUs, at most 16 ones across two nodes.

\begin{figure}[t]
    \centering
    \begin{minipage}[t]{0.5\textwidth} 
        \centering
        \includegraphics[width=\textwidth]{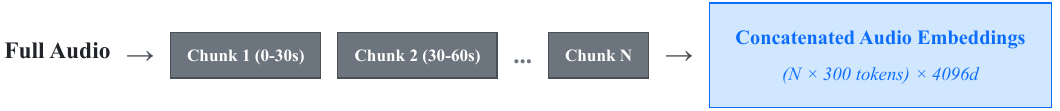}
        \caption{Handing for Full Song Length Audio
        }
        \label{fig:example}
    \end{minipage}%
    \begin{minipage}[t]{0.5\textwidth} 
    \end{minipage}
\end{figure}

\begin{figure*}[t] 
    \centering
    \includegraphics[width=\textwidth]{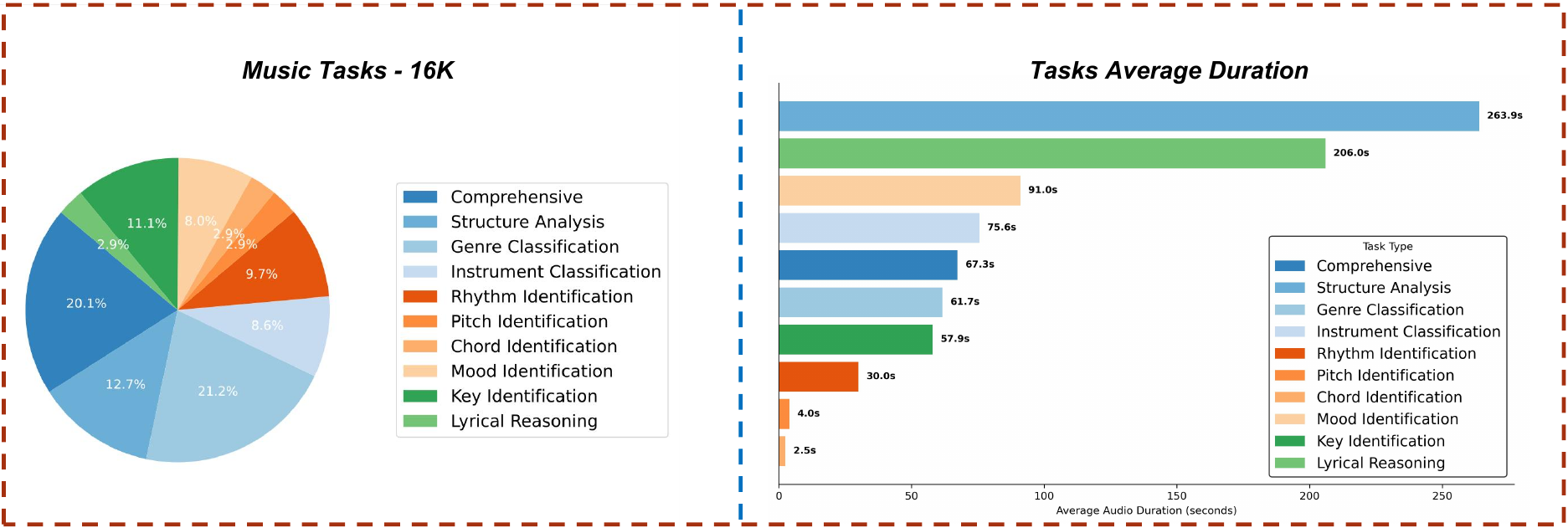} 
    \caption{Dataset distribution of MuCUE. MuCUE contains a total of 16,463 samples. The left side shows the distribution of sample quantities across 10 task categories, while the right side displays the distribution of audio durations among the 10 task categories.}
    \label{fig:example}
\end{figure*}

\subsection{Pre-training}
The objective of the pre-training phase is to establish a strong, fundamental alignment between the audio and language modalities and to imbue the model with core perceptual abilities. This is accomplished over four distinct stages:

\paragraph{Stage 1: Warmup (Connector Initialization)}
The training begins with a brief 400-step warmup stage. In this initial step, we freeze the parameters of both the audio encoder and the LLM, training only the connector module. The rationale is to establish a stable bridge between the two powerful, pre-trained backbones without the risk of destabilizing their well-formed representations with large, chaotic gradients from a randomly initialized component. The tasks are limited to basic speech and music score transcription on 30-second clips, providing a clean, direct signal for the connector to learn the initial modality mapping.

\paragraph{Stage 2: Initial Full-Parameter Alignment (Align1)}
Once the connector is stabilized, we unfreeze all model parameters and begin end-to-end training for 2500 steps. By co-adapting the entire model on the same foundational transcription tasks, we allow the audio encoder and the LLM to gently adjust to one another, deepening the cross-modal alignment beyond the connector. This stage encourages the model to learn a shared representational space more profoundly.

\paragraph{Stage 3: Capability Enrichment (Align2)}
With the model architecture fully aligned, we broaden its musical knowledge base over 3500 steps. The task set is expanded to include more complex and musically salient objectives: lyrics transcription, pitch identification, and instrument identification. This strategic shift moves the model's learning from simple acoustic-to-text mapping towards genuine musical concept recognition. It is at this stage that the model learns the specific acoustic signatures corresponding to textual concepts like "guitar," "C4 pitch," or lyrical content.

\paragraph{Stage 4: Long-Context Extension}
A pivotal and defining stage of our pre-training is the extension to long-form audio. In these 540 steps, the training audio duration is dramatically increased from 30 seconds to a max of 390 seconds. To accommodate the significant memory demands of these long sequences, the batch size is necessarily reduced from 384 to 144. The tasks are focused on music score and lyrics transcription over these extended contexts. The purpose of this stage is to force the model to learn long-range temporal dependencies that are invisible in short clips, such as the verse-chorus structure of a song or the development of a melodic theme. This endows our model with the capacity for true song-level analysis, a critical differentiator from prior work.

\subsection{Fine-tuning}
Following the comprehensive pre-training phase, the model has developed a robust and general understanding of musical concepts. The fine-tuning phase aims to adapt this generalist foundation into a highly proficient expert for a wide array of MIR applications. This is accomplished through a sequential, two-stage fine-tuning curriculum that further refines the model's capabilities, first by mastering diverse tasks on short audio segments and then by applying this knowledge to the complexity of full-length musical pieces.

\paragraph{Stage 1: Short Music Fine-tuning}
The first stage of fine-tuning involves training the model for 1000 steps on a diverse mixture of MIR tasks using 30-second audio segments. The objective here is to expose the model to the rich variety of music understanding tasks found in the wild—from genre and mood classification to more technical analyses like key and tempo detection. By focusing on short, information-dense clips, we can efficiently train the model across this wide task distribution, sharpening its ability to perform fine-grained, segment-level analysis. This stage solidifies the model's grasp of specific musical attributes before it is asked to integrate them over longer time horizons.

\paragraph{Stage 2: Long Music Fine-tuning}
Building directly upon the refined capabilities from the previous stage, the model then undergoes a final 700 steps of fine-tuning, this time using the full 390-second audio context. The purpose of this stage is to transfer the task-specific knowledge learned on short clips to scenarios requiring holistic, song-level reasoning. The model learns to apply its understanding of genre, mood, and structure to entire musical narratives, tracking their evolution and interplay throughout a complete song. This sequential process—first mastering the concepts, then applying them at scale—ensures that the model develops a cohesive and hierarchical understanding, making it adept at both microscopic precision and macroscopic interpretation. This final step is crucial for equipping the model with the comprehensive analytical skills evaluated in our MuCUE benchmark.

\section{Evaluation}
\label{sec:evaluation}
\subsection{The MuCUE Benchmark}

\begin{table*}[t]
    \centering
    \caption{Evaluation Results on various music understanding tasks. We group the datasets by their task category for better readability. The highest score in each row is highlighted in \textbf{bold}.}
    \label{tab:eval-results}
    \resizebox{\textwidth}{!}{%
    \begin{tabular}{@{}ll rrrrr@{}}
        \toprule
        Task & Dataset & Gemini-2.0-flash & Qwen2.5-Omni & Kimi-Audio & Qwen2-Audio & ours \\ \midrule
        \multirow{2}{*}{Key Identification} 
                                            & \texttt{gs\_key\_30s}\cite{Knees2015TwoDS} & 33.6 & 23.8 & 26.0 & 18.2 & \textbf{50.4} \\
                                            & \texttt{gtzan\_key}\cite{1021072} & 33.7 & 28.7 & 28.3 & 22.0 & \textbf{34.1} \\ \midrule
        Pitch Identification & \texttt{nsyn\_pitch}\cite{engel2017neural} & 30.8 & 36.8 & 31.8 & 31.2 & \textbf{77.2} \\ \midrule
        Chord Identification & \texttt{guitarset}\cite{Xi2018GuitarSetAD} & 25.2 & 13.2 & 27.2 & 19.2 & \textbf{58.8} \\ \midrule
        \multirow{2}{*}{Rhythm Identification} & \texttt{ballroom\_tempo}\cite{49348b938edd4f4c9f6f061e216c7828} & \textbf{31.7} & 28.9 & 24.4 & 31.1 & 29.4 \\
                                               & \texttt{gtzan\_tempo}\cite{1021072} & \textbf{41.3} & 32.4 & 22.9 & 27.1 & 40.7 \\ \midrule
        \multirow{3}{*}{Instrument Classification} & \texttt{ins\_cls}\cite{music_instrument_sounds} & 26.0 & 66.8 & 79.4 & 39.8 & \textbf{91.2} \\
                                                   & \texttt{nsyn\_ins}\cite{engel2017neural} & 32.4 & 40.6 & 44.4 & 22.4 & \textbf{74.0} \\
                                                   & \texttt{mtg\_ins}\cite{bogdanov2019mtg} & 19.8 & 55.8 & 51.2 & 24.0 & \textbf{68.6} \\ \midrule
        \multirow{5}{*}{Genre Classification} & \texttt{gtzan}\cite{1021072} & 72.2 & \textbf{88.6} & 77.8 & 83.9 & 81.3 \\
                                              & \texttt{fma-small}\cite{fma_dataset} & 63.4 & 66.2 & 55.8 & 65.6 & \textbf{72.4} \\
                                              & \texttt{fma-medium}\cite{fma_dataset} & 62.8 & 78.0 & 59.8 & 77.0 & \textbf{85.2} \\
                                              & \texttt{mtg\_genre}\cite{bogdanov2019mtg} & 57.2 & 61.6 & 55.8 & 46.4 & \textbf{81.4} \\
                                              & \texttt{ballroom\_genres}\cite{49348b938edd4f4c9f6f061e216c7828} & \textbf{57.0} & 45.8 & 44.0 & 35.2 & 52.4 \\ \midrule
        \multirow{2}{*}{Mood Identification} & \texttt{mtg\_mood}\cite{bogdanov2019mtg} & 38.2 & 43.4 & 39.4 & 29.2 & \textbf{52.8} \\
                                             & \texttt{md4q}\cite{panda2018musical} & \textbf{71.9} & 47.6 & 61.3 & 57.8 & 65.9 \\ \midrule
        \multirow{4}{*}{Structure Analysis} & \texttt{salami\_segd}\cite{smith2011design} & 40.6 & 18.6 & 27.2 & 19.4 & \textbf{64.8} \\
                                            & \texttt{salami\_pred}\cite{smith2011design} & 37.6 & 32.2 & 34.6 & 31.2 & \textbf{64.8} \\
                                            & \texttt{salami\_cnt}\cite{smith2011design} & \textbf{49.8} & 36.8 & 37.8 & 30.2 & 43.2 \\
                                            & \texttt{salami\_overall}\cite{smith2011design} & \textbf{62.1} & 55.8 & 45.3 & 42.6 & 48.7 \\ \midrule
        Lyrical Reasoning & \texttt{lyr}(internal testset for lyrical reasoning) & 88.2 & 87.4 & 87.0 & 60.0 & \textbf{90.8} \\ \midrule
        \multirow{5}{*}{Comprehensive} & \texttt{mmau-music}\cite{sakshi2024mmaumassivemultitaskaudio} & \textbf{67.1} & 63.8 & 66.2 & 57.8 & 66.5 \\
                                       & \texttt{tat}\cite{law2009evaluation} & 61.2 & 59.4 & 54.0 & 61.4 & \textbf{80.6} \\
                                       & \texttt{mucho}\cite{weck2024muchomusic} & 69.6 & 66.5 & \textbf{69.7} & 66.7 & 63.9 \\
                                       & \texttt{mcaps}\cite{agostinelli2023musiclm} & 62.2 & 65.6 & 68.0 & 74.0 & \textbf{80.0} \\
                                       & \texttt{mqa}(internal testset for music QA) & 58.0 & 76.0 & 79.0 & 60.8 & \textbf{88.4} \\ \midrule
        \multicolumn{2}{@{}l}{Average} & 49.8 & 50.8 & 49.9 & 43.6 & \textbf{65.7} \\ \bottomrule
    \end{tabular}%
    }
\end{table*}
The evaluation of large audio models' deep musical intelligence remains a significant challenge, largely due to the lack of a unified and multi-faceted benchmark. To bridge this critical gap, we present MuCUE (Music Comprehensive Understanding Evaluation), a novel benchmark that systematically assesses a wide array of music perception and cognition tasks. Its core innovation lies in framing all evaluation tasks—from low-level pitch and chord recognition to high-level genre, mood, and structural analysis—as multiple-choice questions (MCQs). This standardized format facilitates straightforward, scalable evaluation and is particularly suited for probing the emergent reasoning abilities of generative and foundation models. MuCUE thereby provides the research community with a holistic and rigorous tool to measure the true progress of music AI, identify model weaknesses, and guide future innovations in the field.

The construction process of evaluation data are detailed in appendix.
For datasets in large quantity we hold out a portion for evaluation in the first place and leave the rest for training, so there is no possibility of contamination at least for our model.

\subsection{Main Results and Analysis}

The comprehensive evaluation results on the MuCUE benchmark are presented in Table \ref{tab:eval-results}. We evaluate some open-weighted models\cite{Qwen2-Audio,xu2025qwen2,kimiteam2025kimiaudiotechnicalreport} as well as a proprietary model. 
Upon tests, some other audio language models like MU-LLaMA\cite{liu2023music} have difficulty in doing this kind of MCQs, it becomes unfair to compare in this way, so we exclude them here.
The results unequivocally demonstrate the superior performance of our proposed model. Achieving an average score of 65.7, our model establishes a new state-of-the-art, outperforming the next-best model, Qwen2.5-Omni, by a significant margin of over 15 points in average accuracy. This substantial improvement across a diverse set of 26 tasks underscores the efficacy of our unified architecture and targeted training strategy.

Our model's strength is particularly evident in low-level perceptual tasks that require fine-grained audio analysis. For instance, it achieves remarkable scores of 77.2 on pitch identification (\texttt{nsyn\_pitch}), 58.8 on chord recognition (\texttt{guitarset}), and a commanding 91.2 on instrument classification (\texttt{ins\_cls}). We attribute this success to our architectural design, specifically the fusion of features from multiple encoder layers (0, 7, 15, and 32), which provides the model with a rich, multi-resolution representation of the audio signal. Furthermore, the inclusion of similar tasks during the pre-training alignment phase directly cultivated these foundational capabilities, enabling the model to excel where others falter.

Beyond granular perception, our model excels at high-level cognitive tasks that depend on understanding long-range temporal dependencies. Its exceptional performance on music structure analysis, such as segment boundary detection (\texttt{salami\_segd} at 64.8), and lyrical reasoning (\texttt{lyr} at 90.8), highlights this capability. This proficiency is a direct outcome of our novel context-extending training stage, where the model was explicitly trained on audio contexts of up to 390 seconds. This stage enabled the model to learn the hierarchical and narrative structures inherent in complete musical pieces, a feat unattainable by models limited to short 30-second clips.

A nuanced analysis also reveals areas for further investigation. While dominant in most tasks, our model's performance on overall structural summary (\texttt{salami\_overall} at 48.7) is surpassed by Gemini. This may suggest that tasks requiring high-level abstract summarization are more heavily influenced by the raw reasoning power of the underlying LLM, where proprietary, larger-scale models may hold an advantage. Conversely, on the popular GTZAN genre classification task, our model (81.3) is competitive but behind Qwen2.5-Omni (88.6). This could be attributed to differences in pre-training data composition, as GTZAN is a ubiquitous benchmark that may feature more prominently in the training of other general-purpose models. These results highlight the intricate interplay between model architecture, training data, and task-specific requirements, providing valuable insights for future work.
\subsection{Downstream Applications}
Our trained model is more like a base model, which can be further finetuned to adapt to more specific downstream tasks. For instance, we further train an instruct version and try out reinforcement learning like GRPO(\cite{shao2024deepseekmath}) afterwards. We also make a prompt generator for a music generation model ACE-Step\cite{gong2025acestep} which takes in a song and output some prompts and lyrics for generation. More examples are shown in appendix. This demonstrates the model's strong generalization ability and flexibility.
\subsection{Ablation Study}
\begin{table}[]
\resizebox{\columnwidth}{!}{%
\begin{tabular}{@{}llll@{}}
\toprule
variants                         & instruct & combined finetuning & last hidden \\ \midrule
Avgerage on MuCUE(compared with base) & +0.078   & -0.143             & -1.250      \\ \bottomrule
\end{tabular}%
}
\caption{Ablation Study Results}
\label{tab:ablation}
\end{table}

To validate our key design choices, we conducted several ablation experiments (see Table \ref{tab:ablation}). First, we assess our multi-layer feature fusion by training a variant using only the last hidden layer of the audio encoder with the same data. This resulted in a performance degradation, confirming that a rich, multi-resolution audio representation is critical for our model's success. Next, we test our sequential fine-tuning curriculum by combining the short and long-context data into a single stage. This led to a small performance drop, suggesting our staged approach provides a more effective learning path. Finally, adding a post-hoc instruction-tuning stage yielded a marginal gain, indicating that our core training regimen already aligns the model effectively for question-answering tasks within the music domain. Collectively, these results underscore the importance of our proposed feature fusion and sequential training strategies.
\section{Conclusion and Future Work}
\label{sec:conclusion}

In this work, we introduced a unified foundation model called MuFun that significantly advances the state-of-the-art in holistic music understanding. By leveraging a multi-layer feature fusion architecture and a novel, long-context (390s) training curriculum, our model successfully overcomes the task fragmentation common in Music Information Retrieval. Its superior performance is demonstrated on MuCUE, a comprehensive new benchmark we developed to systematically evaluate a wide spectrum of musical abilities via a standardized multiple-choice question format. Our results validate that a single, strategically trained model can achieve both fine-grained perceptual accuracy and high-level cognitive reasoning, setting a new standard for the field.

While our model demonstrates powerful capabilities, future work can address its current limitations. Our immediate goals are to enhance data efficiency through semi-supervised and self-supervised learning to reduce reliance on large annotated corpora. We also plan to extend the model from a pure understanding system into a unified framework for both music analysis and generation. Further research will focus on developing evaluation methods that capture the subjective and creative aspects of music, and on exploring cross-modal applications that connect music with other domains like video and dance, paving the way for more sophisticated computational creativity.

\bibliography{aaai2026}

\clearpage
\appendix

\section{Model Output Examples}
\begin{figure*}[t] 
    \centering
    \includegraphics[width=\textwidth]{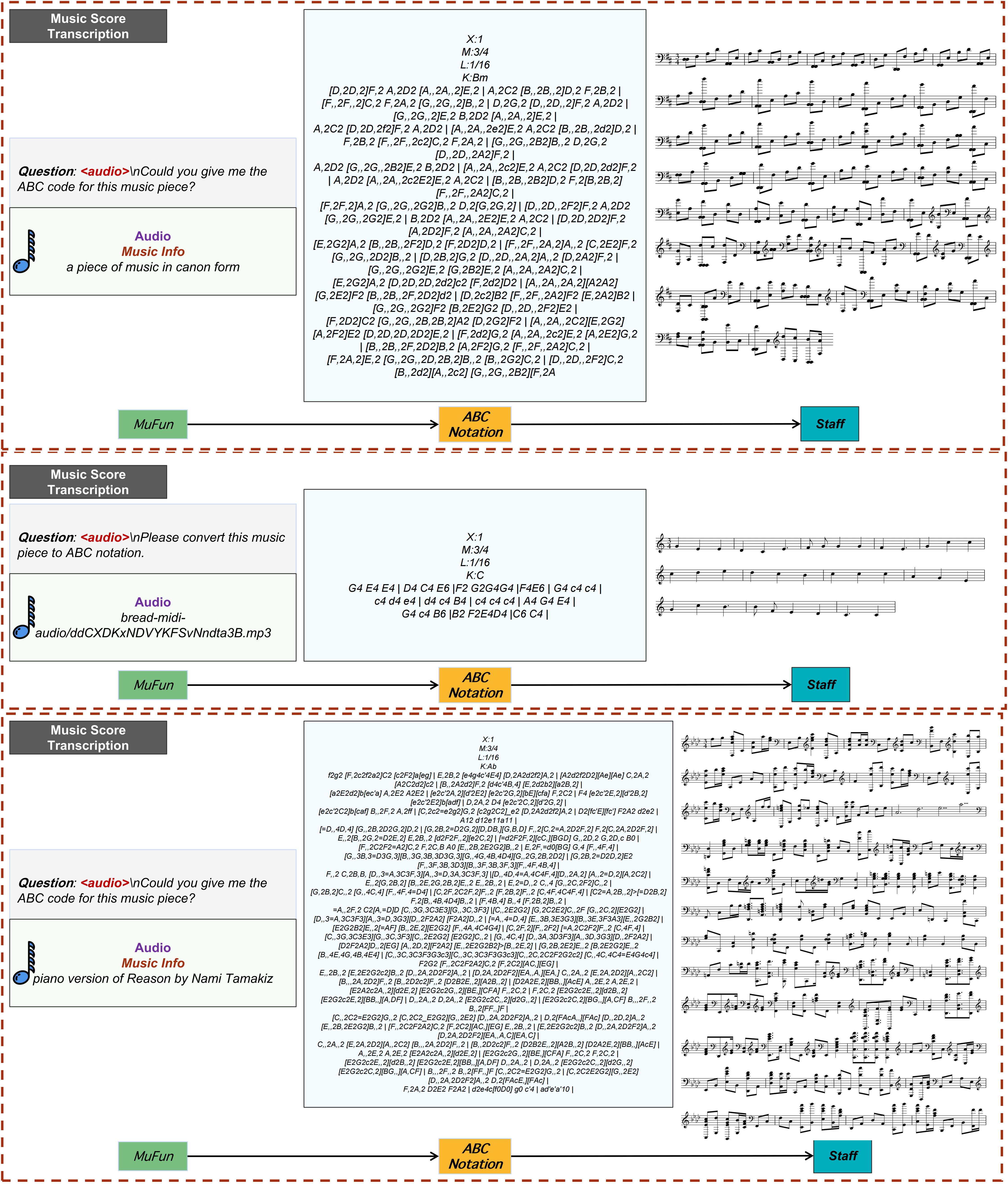} 
    \caption{Music Score Transcription Demo}
    \label{fig:demo9}
\end{figure*}

\begin{figure*}[t] 
    \centering
    \includegraphics[width=\textwidth]{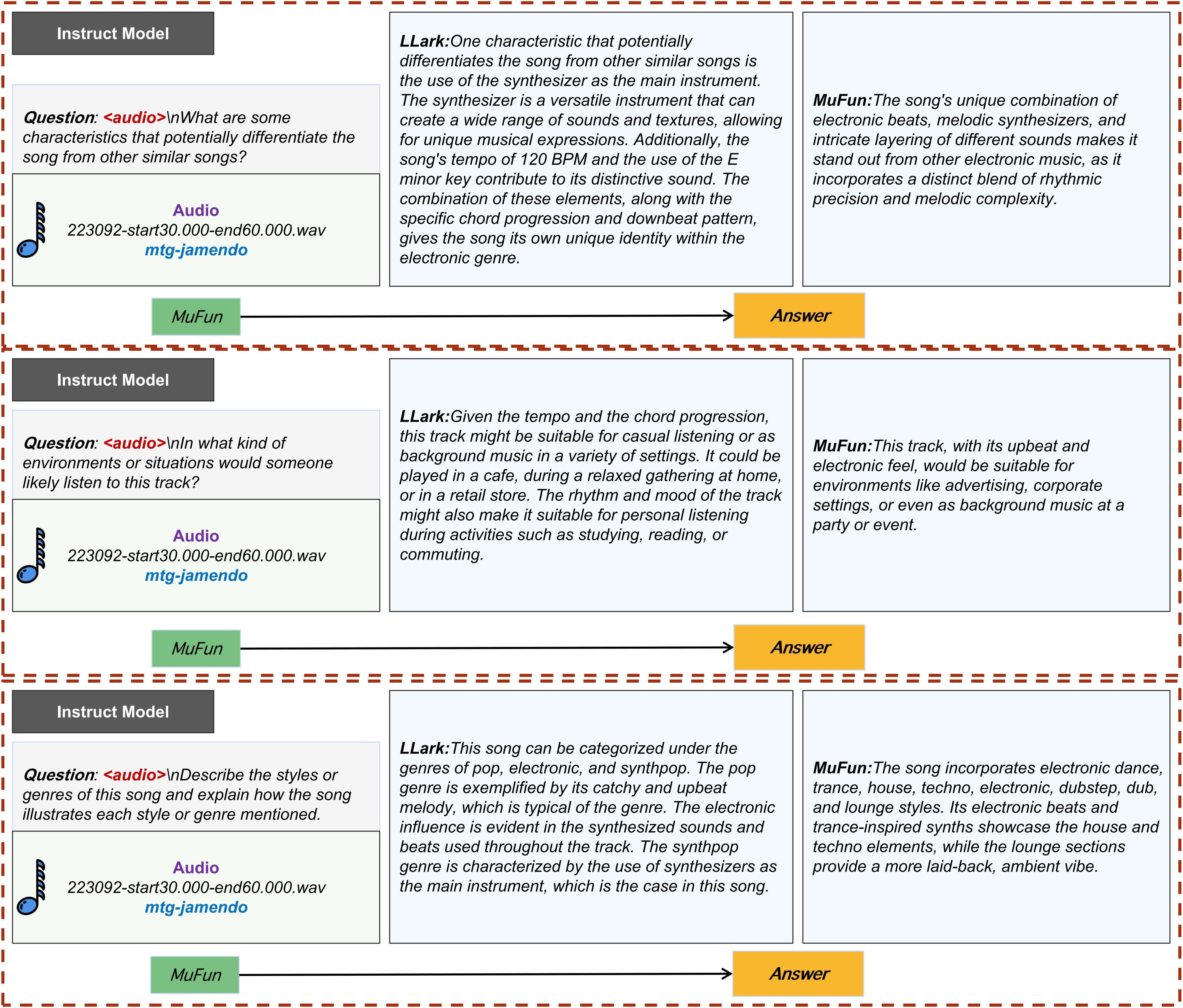} 
    \caption{Instruct Model Demo}
    \label{fig:demo5}
\end{figure*}

\begin{figure*}[t] 
    \centering
    \includegraphics[width=\textwidth]{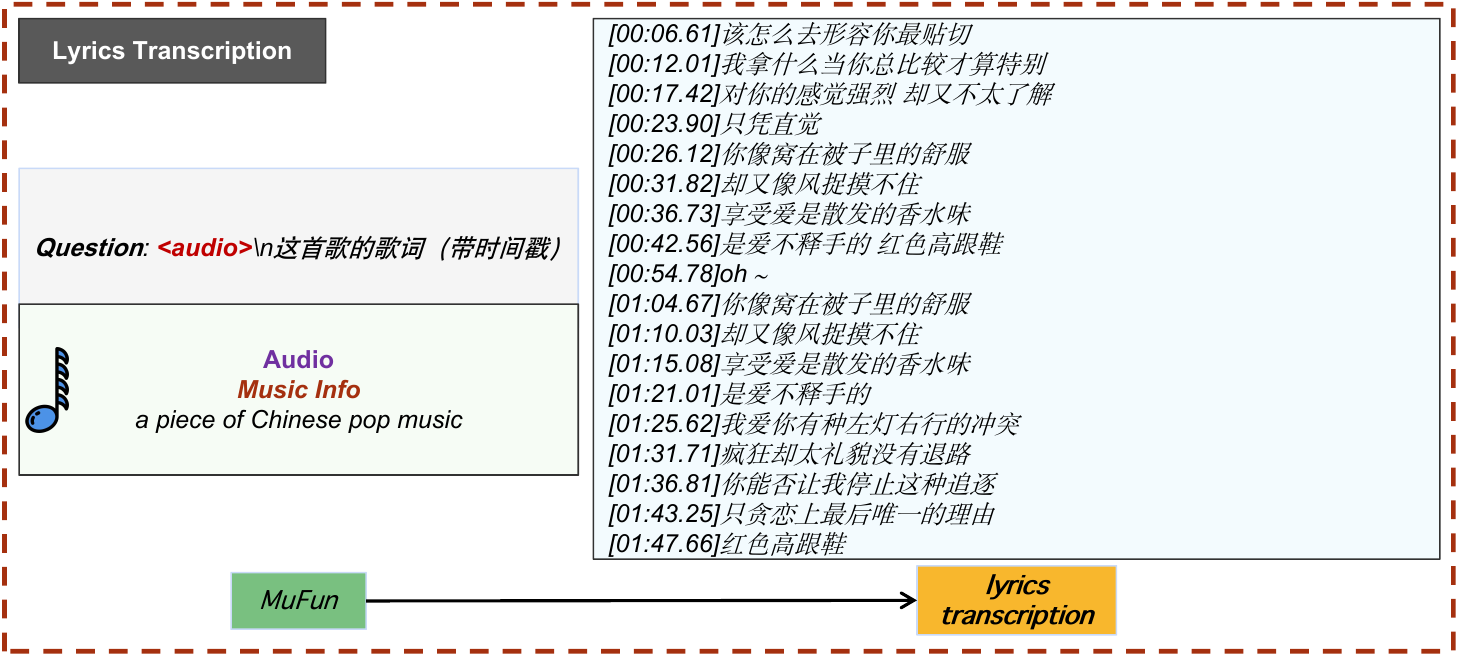} 
    \caption{Lyrics Transcription Demo}
    \label{fig:demo4}
\end{figure*}

\begin{figure*}[t] 
    \centering
    \includegraphics[width=\textwidth]{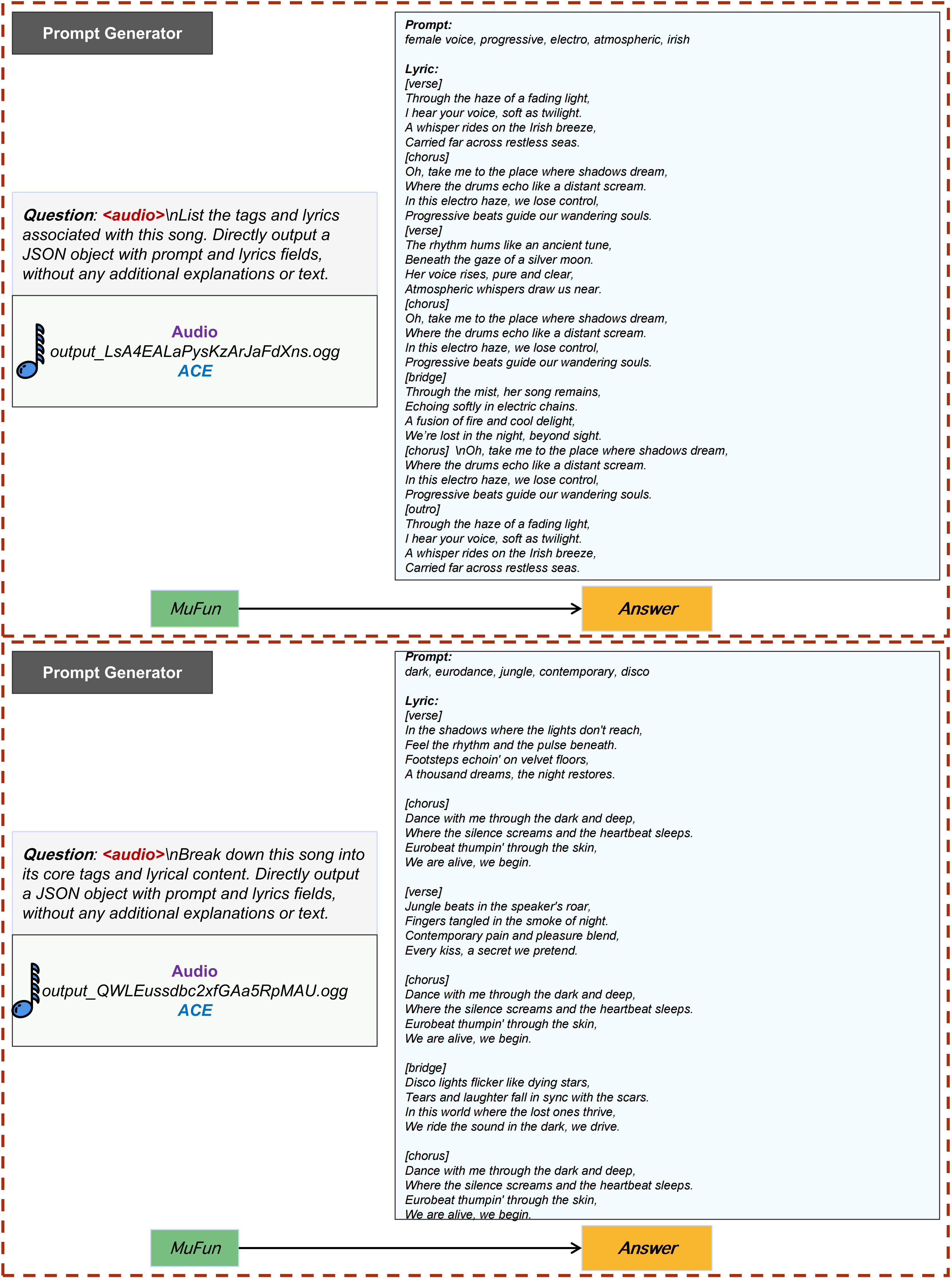} 
    \caption{Prompt Generator Demo}
    \label{fig:demo6}
\end{figure*}

Here we show some outputs of fintunes of our base model, refer to \cref{fig:demo4,fig:demo6,fig:demo5,fig:demo9}.
\section{Training Details}
Our main training code is adapted from TinyLLaVA Factory\cite{zhou2024tinyllava} to support audio input. 
As for reinforcement learning, we modify the HuggingFace TRL library\cite{vonwerra2022trl}. 
\begin{table*}[]
    \resizebox{\textwidth}{!}{%
    \begin{tabular}{@{}llll@{}}
    \toprule
    Dataset                                    & Count  & Audio Length & Labels                                         \\ \midrule
    Giantsteps-key\cite{Knees2015TwoDS}                             & 604    & 120s         & Key                                            \\
    Free Music Archive\cite{fma_dataset}                   & 100k   & 30s          & Similar to website API                         \\
    MagnaTagATune\cite{law2009evaluation}                              & 25k    & 29s          & 188 simple binary tags                         \\
    MTG-Jamendo\cite{bogdanov2019mtg}                                & 55k    & song         & 195 tags (genre, instrument, mood/theme)       \\
    MusicCaps\cite{agostinelli2023musiclm}                                  & 5.5k   & 10s          & Expert-written text descriptions and some tags \\
    YouTube8M-MusicTextClips\cite{mckee2023language}                   & 4k     & 10s          & Text descriptions                              \\
    bread-midi-dataset\cite{matthew_mitton_2025}                         & 851k   & various      & midi                                           \\
    Music Instrument Sounds\cite{music_instrument_sounds} & 42.3k  & 3s           & Instrument                                     \\
    NSynth\cite{engel2017neural}                                     & 306k   & 4s           & Note, instrument                               \\
    GuitarSet\cite{Xi2018GuitarSetAD}                                  & 360    & 30s          & Pitch, beat, tempo, chord                      \\
    Music4All\cite{9145170}                                  & 109k & 30s          & tags, lyrics                                   \\
    SALAMI\cite{smith2011design}                                     & 1447   & song         & structure                                      \\ \bottomrule
    \end{tabular}%
    }
    \caption{open music datasets used in training}
    \label{tab:traindata}
    \end{table*}

For distributed training, we use the DeepSpeed ZeRO Stage 3 optimization strategy. 
In each sub-stage, the model is initialized from previous one and trained with a learning rate of 2e-5 and cosine scheduler. To manage memory and enhance performance, the training utilizes bfloat16 mixed-precision, Flash Attention 2, and gradient checkpointing. 

\begin{figure*}[t] 
    \centering
    \includegraphics[width=\textwidth]{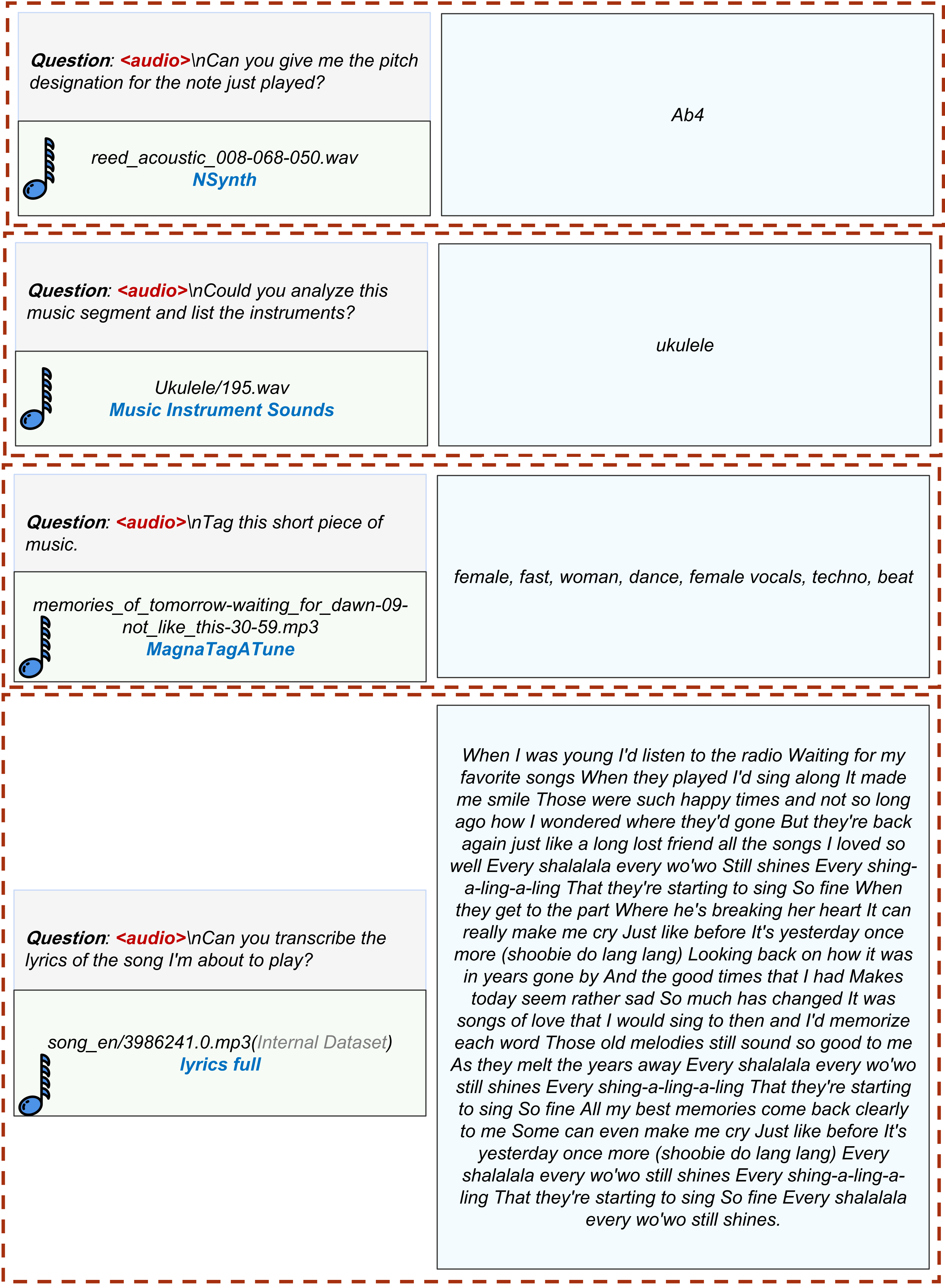} 
    \caption{Training Data Overview
    }
    \label{fig:traind1}
\end{figure*}

\begin{figure*}[t] 
    \centering
    \includegraphics[width=\textwidth]{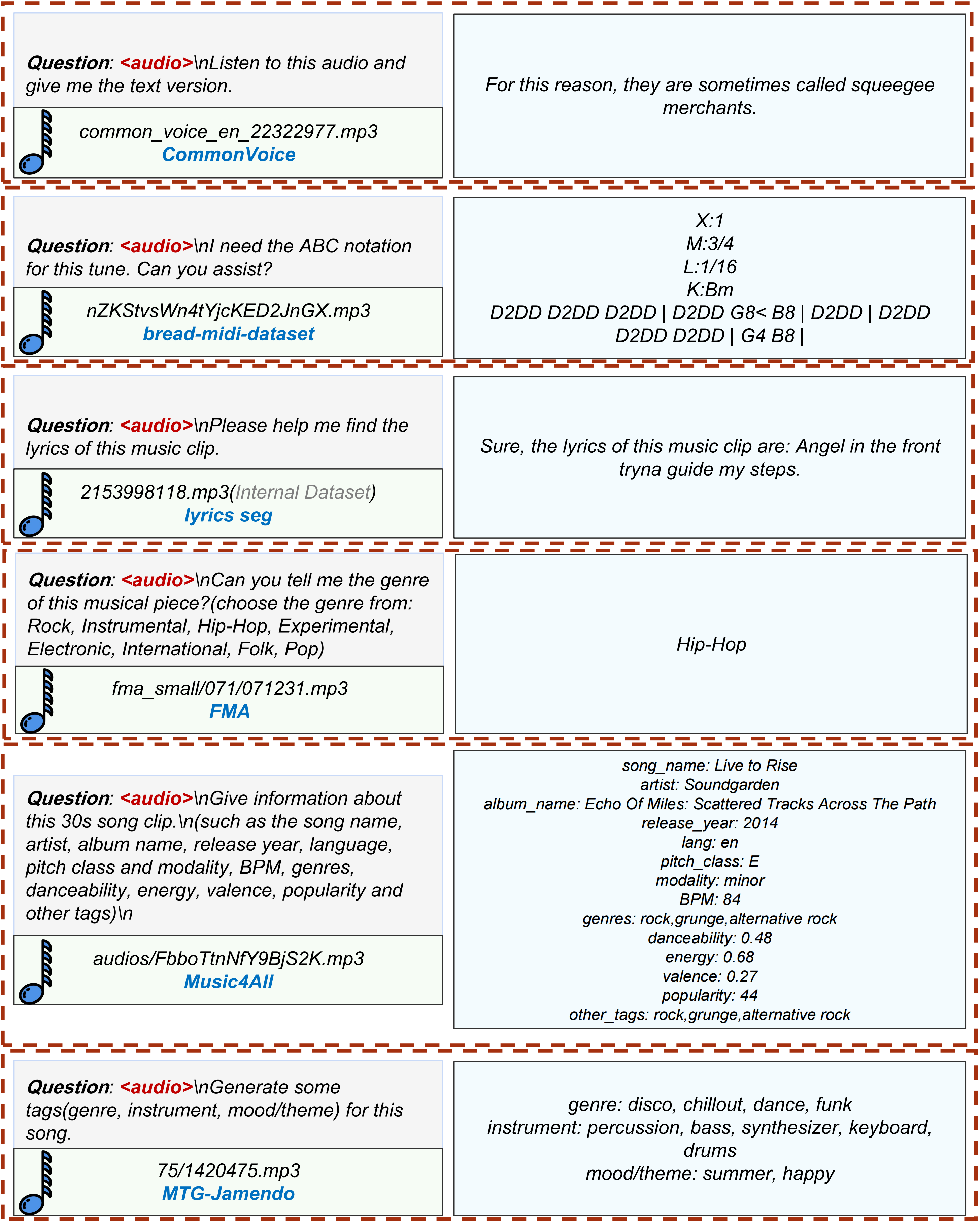} 
    \caption{Training Data Overview
    }
    \label{fig:traind2}
\end{figure*}

The training data is constructed in a straight-forward way, unlike many prior works which mostly utilize instruction format. We tend to put as many text labels as possible in a single sample for better efficiency. Some open music datasets used in training are listed in Table\ref{tab:traindata}, examples shown in \cref{fig:traind1,fig:traind2}. We also provide the specific data recipes for each training stage, see \cref{tab:rwarmup,tab:ralign-1,tab:ralign-2,tab:rcontext-extending,tab:rfinetuning-short,tab:rfinetuning-long}.

\begin{table}[]
\resizebox{\columnwidth}{!}{%
\begin{tabular}{@{}ll@{}}
\toprule
dataset     & nums \\ \midrule
CommonVoice\cite{commonvoice:2020} & 100k \\
bread-midi-dataset\cite{matthew_mitton_2025}       & 38k  \\ \bottomrule
\end{tabular}%
}
\caption{warmup}
\label{tab:rwarmup}
\end{table}

\begin{table}[]
\resizebox{\columnwidth}{!}{%
\begin{tabular}{@{}ll@{}}
\toprule
dataset     & nums  \\ \midrule
CommonVoice\cite{commonvoice:2020} & 1014k \\
bread-midi-dataset\cite{matthew_mitton_2025}       & 115k  \\ \bottomrule
\end{tabular}%
}
\caption{align-1}
\label{tab:ralign-1}
\end{table}

\begin{table}[]
\resizebox{\columnwidth}{!}{%
\begin{tabular}{@{}ll@{}}
\toprule
dataset     & nums \\ \midrule
CommonVoice\cite{commonvoice:2020} & 100k \\
lyrics seg(internal trainset)  & 400k \\
NSynth\cite{engel2017neural}        & 200k \\
Music Instrument Sounds\cite{music_instrument_sounds}         & 42k  \\
People's Speech  \cite{DBLP:journals/corr/abs-2111-09344}       & 310k \\
zhvoice \cite{zhvoice}     & 300k \\ \bottomrule
\end{tabular}%
}
\caption{align-2}
\label{tab:ralign-2}
\end{table}

\begin{table}[]
\resizebox{\columnwidth}{!}{%
\begin{tabular}{@{}ll@{}}
\toprule
dataset     & nums \\ \midrule 
bread-midi-dataset\cite{matthew_mitton_2025}     & 10k  \\
lyrics full(internal trainset)  & 100k \\ \bottomrule
\end{tabular}%
}
\caption{context-extending}
\label{tab:rcontext-extending}
\end{table}

\begin{table}[]
\resizebox{\columnwidth}{!}{%
\begin{tabular}{@{}ll@{}}
\toprule
dataset    & nums \\ \midrule
lyrics seg(internal trainset)  & 4k   \\
MagnaTagATune\cite{law2009evaluation}        & 71k  \\
NSynth\cite{engel2017neural}        & 88k  \\
MusicCaps\cite{agostinelli2023musiclm}         & 10k  \\
Music Instrument Sounds\cite{music_instrument_sounds}       & 52k  \\
GuitarSet\cite{Xi2018GuitarSetAD}    & 4k   \\
 Giantsteps-key\cite{Knees2015TwoDS}      & 3k   \\
FMA-medium\cite{fma_dataset}    & 24k  \\
FMA-small\cite{fma_dataset}      & 7k   \\
Music4All\cite{9145170}        & 109k \\ \bottomrule
\end{tabular}%
}
\caption{finetuning-short}
\label{tab:rfinetuning-short}
\end{table}

\begin{table}[]
\resizebox{\columnwidth}{!}{%
\begin{tabular}{@{}ll@{}}
\toprule
dataset     & nums \\ \midrule
lyrics full(internal trainset)  & 6k   \\
finetuning-short   & 10k  \\
MTG-Jamendo\cite{bogdanov2019mtg}        & 59k  \\
music general QA (internal trainset)        & 10k  \\
lyrical reasoning (internal trainset)         & 9k   \\
SALAMI\cite{smith2011design}       & 5k   \\ \bottomrule
\end{tabular}%
}
\caption{finetuning-long}
\label{tab:rfinetuning-long}
\end{table}
\section{Evaluation Details}
\subsection{Data Construction}
We collect 10+ open datasets as well as some in-house data. 
Since these datasets have audio and text label pairs, multiple-choice questions can be easily generated for evaluation. 
The correct choice can be inferred from the text label pair, while the incorrect choices can be randomly selected from the text label in other pairs.
For example, the fma-small dataset has 8 genre labels, we can randomly select 3 incorrect choices from the other 7 genres.
The exception is the MusicCaps, for which we directly prompt gpt-4o to generate the multiple-choice questions based on the corresponding text descriptions.
For MMAU\cite{sakshi2024mmaumassivemultitaskaudio}, the music part of test-mini v05.15.25 version is used.
\subsection{Experiments}
The input text prompt is 'Choose the correct option for the question based on the audio.', followed by the question and choices. For open-weighted model and our model, we all use BF16 precision for inference and temperature zero during sampling. The gemini model is accessed via Vertex AI Platform with default request parameters.

\section{Downstream Applications}

\subsection{Instruction-tuning}
The instruction-tuning data for full song understanding are scarce, so we prompt various LLMs to generate some QA pairs based on some text information related to a song. This dataset has around 30k samples with all song-level audios, on which the base model is further trained.  Afterwards, we also try methods of reinforcement learning like GRPO(\cite{shao2024deepseekmath}) on it.
\subsection{Prompt Generator}
 We make a prompt generator for a music generation model ACE-Step\cite{gong2025acestep} which takes in a song and output some prompts and lyrics for generation, by training on around 20k text song pairs.
 On a held-out test set of 100 songs, our finetuned model achieves a similarity score of 0.98 (using MERT\cite{li2023mert} model's embedding) compared with original generated song.
\subsection{Music Score Transcription}
The model is trained on more instrumental piece ABC code pairs to have better performance on music score transcription task. ABC code can be used to synthesized back the music. On a held-out test set of size 256, our finetuned model achieves a similarity score of 0.92 (using MERT's embedding) compared with original piece.
\subsection{Lyrics Transcription with Timestamps}
We also try to train a model that could output both lyrics and timestamps given a song.

\end{document}